\documentclass[floatfix,aps,twocolumn,preprintnumbers,amsmath,amssymb,nofootinbib,superscriptaddress,showkeys,showpacs]{revtex4-2}
\usepackage{amsmath}
\usepackage{graphicx}
\usepackage{hyperref}
\usepackage{natbib}
\usepackage[final]{changes}

\newcommand{\nh}{{\bf h}}

\newcommand{\np}{{\bf p}}
\newcommand{\nq}{{\bf q}}

\begin{document}
\title{
\deleted{SuSAM-v2: } \replaced{I}{i}mproved superscaling description 
of electron and \replaced{charged-current}{CC} neutrino quasielastic scattering 
using effective mass dynamics
}

\author{V.L. Martinez-Consentino}\email[]{victormc@ugr.es}
\affiliation{Departamento Sistemas F\'isicos Qu\'imicos y Naturales,
  Universidad Pablo de Olavide, Sevilla, E-41013, Spain.}
\affiliation{Departmento de Ciencias Integradas, Universidad de
  Huelva, E-21071 Huelva, Spain.}

\author{P.R. Casale} \email{palomacasale@ugr.es}
\affiliation{Departamento de
  F\'{\i}sica At\'omica, Molecular y Nuclear
 Universidad de Granada,
  E-18071 Granada, Spain.}
\affiliation{Instituto Carlos I
  de F{\'\i}sica Te\'orica y Computacional,
 Universidad de Granada,
  E-18071 Granada, Spain.}

\author{J.E. Amaro}\email{amaro@ugr.es} 
\affiliation{Departamento de
  F\'{\i}sica At\'omica, Molecular y Nuclear
 Universidad de Granada,
  E-18071 Granada, Spain.}
\affiliation{Instituto Carlos I
  de F{\'\i}sica Te\'orica y Computacional,
 Universidad de Granada,
  E-18071 Granada, Spain.}


\begin{abstract}
We present an improved version of the Superscaling Analysis with
Relativistic Effective Mass, denoted as SuSAM-v2. In the
original SuSAM model, a universal scaling function was fitted to a
selected set of quasielastic electron scattering \((e,e')\) cross
section data, using a phenomenological ansatz inspired by the
Relativistic Mean Field model of nuclear matter. In this work,
we refine the procedure by first fitting a longitudinal scaling
function directly to experimental longitudinal response
data. Subsequently, a separate transverse scaling function is
extracted from purely transverse data, after subtracting the
longitudinal contribution already determined. We find that the
resulting transverse scaling function must exhibit an explicit
dependence on the momentum transfer \(q\) in order to reproduce all
kinematics consistently. The resulting SuSAM-v2 model simultaneously 
describes inclusive quasielastic cross sections and both longitudinal
and transverse response functions in electron scattering. The model is
then applied to neutrino-nucleus scattering, showing an improved
prediction compared to the previous SuSAM-v1 version, due to a more accurate
treatment of the relative contributions of the longitudinal and
transverse weak nuclear responses.
\end{abstract}

\maketitle

\section{Introduction}

Electron and neutrino scattering from nuclei play a central role in
advancing our understanding of nuclear dynamics and in the
interpretation of neutrino oscillation experiments
\cite{Don84,Ben08}. In the case of electron scattering, precise
measurements provide access to the charge and current distributions
involved in nuclear transitions \cite{Jou96}. For neutrino
interactions, a detailed knowledge of neutrino–nucleus cross sections
is essential for reconstructing the incident neutrino energy in
oscillation analyses \cite{For12,Alv18}. However, modeling
lepton–nucleus interactions is challenging due to the complexity of
nuclear effects. Approaches such as the Relativistic Fermi Gas
(RFG) provides a simple baseline description of the quasielastic (QE)
cross section \cite{Smi72}, but they lack important features such as
medium modifications, finite-size effects, nuclear correlations, and
final-state interactions. More refined nuclear models are therefore
required to achieve an accurate and reliable description of the data.

A wealth of studies have been devoted to modeling the intricate nuclear
dynamics underlying electron and neutrino scattering. Among these are
relativistic distorted-wave impulse approximation models based
on the nuclear shell structure \cite{Udi99,Cab05,Gon20}, continuum
random phase approximation approaches that include long-range
correlations \cite{Meu11,Nie11,Pan14}, and spectral function models
that provide a detailed account of single-particle motion and
nucleon–nucleon correlations \cite{Ben89,Ben08}. Additional
sophistication is introduced through the inclusion of meson-exchange
currents (MEC) \cite{Ama03,Ama09,Fra23}, short-range (SRC) and tensor
correlations \cite{Fab97}, and final-state
interactions \cite{Pet02}. Moreover, {\em ab initio} methods such as Green’s
function Monte Carlo  are increasingly able to describe light-
and medium-mass nuclei with high precision
\cite{Car02,Car15,Lov14}. While these approaches yield valuable
insights, they are computationally demanding and often limited in
scope when applied to the full range of kinematics or nuclear targets
needed in neutrino oscillation experiments.  Moreover, despite the
level of sophistication achieved in these theoretical approaches, they
often lead to different predictions for the cross section or response
functions under comparable kinematic conditions, making it difficult
to establish a unique and reliable description of the data.

In this context, simpler yet accurate frameworks that capture the
dominant physics across many nuclei and kinematics remain essential.
Models based on superscaling provide semi-phenomenological
parametrizations of the inclusive quasielastic cross section
\cite{Don99}. They typically assume a factorization between an average
single-nucleon cross section and a function that encapsulates the
effective number of nucleons that can be excited by transferring
momentum $q$ and energy $\omega$ \cite{Alb88,Bar98}. This function,
closely related to the integrated hole-spectral function, represents
the distribution of available nucleons capable of undergoing a
transition within the energy interval $[\omega, \omega + d\omega]$. By
dividing out the known RFG dependence on $q$ and $k_F$, and the
single-nucleon dynamics, one arrives at a universal scaling
function, $f(\psi)$, that ideally depends only on a single dimensionless variable
$\psi$ characterizing the kinematics of the one-body nuclear response
\cite{Day02}.

In the original superscaling analysis (SuSA) approach, the
phenomenological longitudinal scaling function $f_L(\psi)$ was
extracted from selected quasielastic longitudinal response data
\cite{Don99, Mai02}. It was then assumed that the transverse scaling
function $f_T(\psi)$ could be approximated by the same function,
$f_T(\psi) \approx f_L(\psi)$, in accordance with the idea of zeroth-kind 
scaling \cite{Don99b}. However, this assumption proved to
be inconsistent with experimental data, where a transverse enhancement
is significant \cite{Bod22,Bod24,Bod24b}.
In the updated SuSA-v2 framework, the transverse
response was instead modeled using results from the relativistic mean
field (RMF) approach, which better captured the observed behavior
\cite{Gon14,Ama21}. This improvement required introducing a mild but
necessary $q$-dependence in the transverse scaling function, thereby
explicitly breaking superscaling in that channel while preserving
agreement with data.

In this work, we build upon the superscaling analysis with
relativistic effective mass (SuSAM), a variant of SuSA designed to
improve the description of quasielastic $(e,e')$ data by incorporating
nuclear dynamics inspired by the RMF theory of nuclear matter
\cite{Mar17,Ama18}.  The motivation for using RMF-inspired dynamics
lies in its long-standing success in describing the shape and
magnitude of the quasielastic response \cite{Ros80,Weh93}.  Unlike the
original SuSA approach, which is based on the RFG with a fixed nucleon
mass and a phenomenological energy shift, SuSAM introduces a dynamical
ingredient at the single-nucleon level through the use of a
relativistic effective mass $M^*$, as derived from the Walecka model
\cite{Hor81,Ser86}. This approach embeds medium effects from the
outset and preserves gauge invariance, which is not guaranteed in the
traditional SuSA prescription. Another key difference is the fitting
strategy: while SuSA derives the scaling function from selected
longitudinal response data, SuSAM directly fits a phenomenological
scaling function to selected quasielastic cross section data.  This
strategy allows for a global analysis of inclusive electron scattering
data and has been successfully used to extract the effective mass
$M^*$ and Fermi momentum $k_F$ for a wide range of nuclei across the
periodic table \cite{Ama18}. The model has also demonstrated good
predictive power for neutrino-nucleus cross sections, particularly
after including contributions from two-particle–two-hole (2p2h)
excitations computed consistently within the same RMF framework
\cite{Mar21b,Mar23a,Mar23b}.

Despite the success of SuSAM in describing inclusive quasielastic
cross sections, it falls short when it comes to reproducing the
individual longitudinal and transverse nuclear responses. This
limitation arises because the model was directly fitted to inclusive
$(e,e')$ cross section data, which effectively represent a weighted
average of the longitudinal and transverse components. As a result,
SuSAM does not provide sufficient control over each response
separately, especially at low momentum transfer. This issue becomes
particularly relevant for neutrino-nucleus scattering, where the
longitudinal and transverse responses enter the cross section in a
different combination than in electron scattering, and where
additional contributions from the axial current also play a role.

In this work, we propose an improved scaling analysis—SuSAM-v2—that
aims to overcome this limitation. The model builds upon the SuSAM
framework by preserving its key dynamical ingredient, the relativistic
effective mass, while introducing a more constrained and controlled
extraction of the scaling functions. Specifically, we fit the
longitudinal scaling function to data on the separated longitudinal
response and simultaneously to inclusive $(e,e')$ cross sections. The
transverse response is not fitted directly but emerges as a prediction
of the model. Remarkably, this approach leads to a good description of
all available data, including both separated response functions and
inclusive cross sections, thus providing a more robust foundation for
neutrino cross section predictions.

As in the case of SuSA-v2, this strategy entails introducing a
transverse scaling function that exhibits an explicit dependence on
the momentum transfer $q$. Nevertheless, this $q$-dependence can be
captured through a simple analytical parameterization, which maintains
the overall transparency and usability of the model.
\added{Superscaling-based models such as SuSAM$^*$v2 and SuSA-v2 rely
on a phenomenological analysis aimed at reproducing experimental data
with minimal theoretical assumptions. This approach enhances their 
utility in practical applications, such as neutrino event generators, 
but also highlights the importance of recognizing their intrinsic 
limitations, particularly concerning the microscopic nuclear dynamics
that are not explicitly modeled.}

The work is organized as follows: in Sec. \ref{seccion2}, we present a
summary of the formalism for electron scattering, including the cross
section structure, the RMF model for nuclear
matter, and the superscaling approach incorporating an effective
nucleon mass. In Sec. \ref{seccion3}, we outline the development of
the SuSAM-v2 model, describing the modifications introduced to improve
the separate description of the nuclear response functions. In
Sec. \ref{seccion4}, we compare the predictions of SuSAM-v2 with
experimental data from both electron and CC neutrino scattering.
Finally, Sec. \ref{seccion5} contains our conclusions.

\section{Formalism}
\label{seccion2}

The starting point for the scaling analysis of the quasielastic
response is traditionally the RFG model, which provides a simple
framework including basic features of the momentum distribution of the
nucleons with Fermi momentum $k_F$, and the one-body interaction with
the lepton.  In the SuSAM approach, this starting point is extended
by incorporating RMF dynamics.  The nucleons are now described as
relativistic plane waves moving in the presence of a (constant)
RMF that includes both scalar and vector
components. The attractive scalar potential lowers the mass of the
nucleons, giving rise to a relativistic effective mass \( m_N^* = M^*
m_N \), with \( M^* < 1 \), typically \( M^* = 0.8 \) for \(
^{12}\mathrm{C} \). The repulsive vector potential shifts the energy
levels upward. As a result, the total energy of a nucleon in the
medium is given by
\[
E_{\text{RMF}} = E + E_v,
\]
where \( E = \sqrt{(m_N^*)^2 + p^2} \) is the on-shell energy
of a nucleon with momentum $\np$, and \( E_v \) is the
vector potential energy. For $^{12}C$, $E_v=141$ MeV (see Ref. ~\cite{Mar21a}).

The $(e,e')$ cross-section in
plane-wave approximation can be written as a linear combination
of  longitudinal and transverse response functions 
\begin{equation}
\frac{d\sigma}{d\Omega'd\epsilon'}
= \sigma_{\rm M}
(v_L R_L (q,\omega)+  v_T  R_T(q,\omega)).
\end{equation}
Here, $\sigma_{M}$ is the Mott
cross-section, $q$ is 
the momentum transfer, $\omega$ the
energy transfer, and  the four-momentum transfer verifies $Q^2=\omega^2-q^2 <0$.
The kinematical factors are
\begin{eqnarray}
v_L &=& 
\frac{Q^4}{q^4} \\
v_T &=&  
\tan^2\frac{\theta}{2}-\frac{Q^2}{2q^2}.
\end{eqnarray}
where $\theta$ is the scattering angle.

In the impulse approximation the
electron is assumed to scatter off individual nucleons inside the
nucleus through the one-body current, inducing one-particle–one-hole
(1p1h) excitations in the nuclear system.
The response functions are then
\begin{eqnarray}
  R_K(q,\omega)
  &=& \frac{V}{(2\pi)^3}\int d^3h
  \frac{(m_N^*)^2}{E_pE_h}
  2 w_K(\np,\nh)
\nonumber\\
&&\times  \delta(\omega + E_h - E_p) 
\theta(p-k_F)\theta(k_F-h)
\label{respuesta}
\end{eqnarray}
where $K=L,T$, and $\np=\nh+\nq$ by momentum conservation.
The functions  $w_K(\np,\nh)$ are the responses for the 1p1h excitation.
With the $z$-axis aligned along the momentum transfer $\nq$ they are
\begin{equation}
  w_L=w^{00}, \kern 1cm w_T=w^{xx}+w^{yy}
\end{equation}
where $w^{\mu\nu}$ is the single-nucleon hadronic tensor
\begin{equation}
  w^{\mu\nu}(\np,\nh)=
  \frac12 \sum_{s_ps_h} j^{\mu}(\np,\nh)^*j^\nu(\np,\nh),
\end{equation}  
and $j^\mu$ is the electromagnetic current of the nucleon
\begin{equation}
  j^\mu(\np,\nh) = \bar{u}_{s_p}(\np)
  \left[ F_1(Q^2)\gamma^\mu +
    F_2(Q^2)\frac{i\sigma^{\mu\nu}Q_\nu}{2m_N} \right] u_{s_h}(\nh),
\end{equation}
where $u_s$ are Dirac spinors with mass $m_N^*$.  The RMF
response described above corresponds to either protons or neutrons
separately; the total nuclear response is then obtained by summing the
proton and neutron contributions.  Note that the vector potential
energy, $E_v$ cancels out in the energy difference between the
particle and hole states, Eq. (\ref{respuesta}), so the nuclear
response does not depend on it.

One of the key assumptions of scaling approaches is the factorization
of the nuclear response into a averaged single-nucleon response and a function
that encapsulates the nuclear dynamics. In the RFG
model, this factorization is exact, as can be readily seen in the
following way. We first integrate over the angular variables in Eq. (\ref{respuesta}), using
the energy-conserving delta function, and arrive at
\begin{equation} \label{respuesta2}
R_K(q,\omega)=
\frac{V}{(2\pi)^3}
\frac{2\pi m_N^*{}^3}{q}
\int_{\epsilon_0}^{\epsilon_F} 
n(\epsilon)
2  w_K(\epsilon,q,\omega).
\end{equation}
Here $\epsilon=E_h/m_N^*$ is the hole energy in units of the effective nucleon
mass, $\epsilon_F=E_F/m_N^*$ refers to the Fermi energy in the
same units and $n(\epsilon)=\theta(\epsilon_F-\epsilon)$. The lower
limit $\epsilon_0$ is given by (see Appendix C of \cite{Ama20} for a
detailed derivation)
\begin{equation}
  \epsilon_0=
 {\rm Max}
 \left\{ \kappa\sqrt{1+\frac{1}{\tau}}-\lambda, \epsilon_F-2\lambda
 \right\}.
\end{equation}
Here we are using dimensionaless variables
\begin{equation}
\kappa=\frac{q}{2m_N^*}, \kern 1cm
\lambda=\frac{\omega}{2m_N^*}, \kern 1cm
\tau= \kappa^2-\lambda^2.
\end{equation}
Now we can define an average value of the single-nucleon response 
 as follows:
\begin{equation}
\overline{w}_K(q,\omega)=
\frac{  \int_{\epsilon_0}^{\epsilon_F} n(\epsilon)  w_K(\epsilon,q,\omega)}
{ \int_{\epsilon_0}^{\epsilon_F}  n(\epsilon)},
\end{equation}
allowing us to express the total response in a factorized form:
\begin{equation} \label{respuesta22}
R_K(q,\omega)=
\frac{V}{(2\pi)^3}
\frac{2\pi m_N^*{}^3}{q}
2\overline{w}_K(q,\omega)
\int_{\epsilon_0}^{\infty} 
n(\epsilon)
\end{equation}
This expression can be further transformed by:
\begin{itemize}
\item  using that the nuclear
volume $V$ is related to the number of nucleons $\mathcal{N}$ through
the Fermi momentum $k_F$ as
$
\frac{V}{(2\pi)^3} = \frac{\mathcal{N}}{\frac{8}{3}\pi k_F^3}$,

\item defining the scaling function $f(\psi)$ as
  \begin{equation}
 \frac34  \int_{\epsilon_0}^{\infty}n(\epsilon)d\epsilon \equiv (\epsilon_F-1) f^*(\psi^*),
\end{equation}
 
\item and definig the scaling variable as
\begin{equation}
\psi^* = \sqrt{\frac{\epsilon_0-1}{\epsilon_F-1}} {\rm sgn} (\lambda-\tau).
\end{equation}
\end{itemize}
The sign convention is chosen so that $\psi^*$ is negative when $\lambda < \tau$ (to the left of the quasielastic peak), equals zero at the peak ($\lambda = \tau$), and is positive when $\lambda > \tau$ (to the right of the peak).

Note that with these definitions, the scaling function takes the simple parabolic form
\begin{equation}
f^*(\psi^*) = \frac{3}{4}(1 - \psi^*{}^2)\theta(1-\psi^*{}^2) \,.
\end{equation}
With this, the nuclear response function can be written as
\begin{equation}
  R_K(q,\omega) =
  \overline{w}_K(q,\omega)
  \mathcal{N}
    \frac{2}{q}\left(\frac{m_N^*}{k_F}\right)^3
  (\epsilon_F-1) f^*(\psi^*) \,.
\end{equation}
This expression can be transformed into the simplified form
\begin{equation}
R_K(q,\omega)  =   r_K(q,\omega) f^*(\psi^*),
\label{factorization}  
\end{equation}
by introducing the so-called single-nucleon prefactors $r_K$,
\begin{equation}
r_K(q,\omega)=
\overline{w}_K(q,\omega)
\mathcal{N}
\frac{2}{q}\left(\frac{m_N^*}{k_F}\right)^3
  (\epsilon_F-1),
\end{equation}
which contain not only the averaged nucleon responses but also other
kinematic and normalization factors that, strictly speaking, belong to
the full nuclear response. Nevertheless, the term “single-nucleon
factors” has become standard in the literature. In any case, the total
response is obtained by combining the proton and neutron
contributions
\begin{equation}
  R_K(q,\omega)  = (Z  r^p_K+N r_K^n)
  f^*(\psi^*),
\label{susam}
\end{equation}
where the prefactors $r_K^p$ ($r_K^n$) contain the averaged response of the
proton (neutron), $\overline{w}^p_K$ ($\overline{w}^n_K$).
When the nuclear response is divided by the total prefactor
$(Z  r^p_K+N r_K^n)$, what
remains is the superscaling function $f^*(\psi^*)$, which is independent of the momentum
transfer $q$ and of the nuclear species when expressed as a function
of the scaling variable $\psi^*$. This variable has been constructed
such that its range always lies within the interval $[-1, 1]$.

In the SuSAM approach, the selected quasielastic $(e,e')$ cross section data are divided by the corresponding combination of longitudinal and transverse single-nucleon prefactors, including the appropriate kinematic factors and the Mott cross section. This defines the phenomenological superscaling function $f^*(\psi^*)$.
\begin{equation}
  f_{\rm exp}^*(\psi^*) =
\frac{ \left( \frac{d\sigma}{d\Omega d\omega} \right)_{\text{exp}} }
     { \sigma_\text{Mott} \left[ v_L (Z r_L^p + N r_L^n) + v_T (Z r_T^p + N r_T^n) \right] }
  \end{equation}

Among the data scaled using the procedure described above, a careful
selection was carried out to isolate genuine quasielastic events. This
was achieved by excluding data points that exhibited large deviations
from the scaling trend, typically associated with inelastic
excitations or very low energy transfers, where scaling violations are
more prominent. This selection strategy was systematically applied in
a series of works, initially focused on ${}^{12}$C, and later extended
to a global fit encompassing nuclei from deuterium up to uranium
\cite{Mar17,Ama18}. The result was a remarkably good overall
description of inclusive quasielastic electron scattering cross
sections across the nuclear chart using a small number of parameters
and a single phenomenological scaling function. This function was
parametrized through a sum of Gaussians to capture its shape
accurately. The model was subsequently applied to neutrino scattering
data \cite{Rui18}, where the inclusion of 
2p2h excitations proved essential \cite{Mar21b,Mar23b}. In
further developments, the model was refined to account for the
high-$\psi^*$ tail of the scaling function. This tail was
interpreted as arising from processes involving the emission of two
nucleons, potentially reflecting the underlying dynamics of
short-range correlated nucleon pairs \cite{Mar23a}.

\section{Development of SuSAM-v2}
\label{seccion3}

Despite the overall success of scaling models based on a single
phenomenological scaling function in describing quasielastic
lepton–nucleus scattering, the SuSAM-v1 framework presents some
limitations. In particular, it does not accurately reproduce
the separated longitudinal ($R_L$) and transverse ($R_T$) nuclear
response functions. The use of an effective nucleon mass within
SuSAM-v1 allows for a successful description of the inclusive cross
section, but only as some average of $R_L$ and $R_T$, rather than
capturing each component correctly. This limitation becomes especially
relevant in the context of neutrino scattering, where the different
nuclear response functions contribute with distinct weights. Moreover,
the transverse channel is significantly enhanced by the presence of
axial-vector currents, which contribute in addition to the purely
vector responses present in electron scattering. There also appears a
new interference response, $R_{T'}$, arising from the coupling between
vector and axial currents. In this section, we construct the SuSAM-v2
model as an improved approach aimed at addressing these shortcomings
by enabling a more accurate and consistent description of the
separated response functions.

A key step in refining the SuSAM-v2 model is the introduction of two
independent scaling functions, $f_L$ and $f_T$, corresponding to the
longitudinal and transverse nuclear responses, respectively. This
extension is motivated by the observation that the longitudinal
response function, $R_L$, approximately exhibits superscaling
behavior, making it suitable for the direct extraction of $f_L$ from
experimental data. In contrast, the transverse response $R_T$ does not
scale as well due to the influence of additional reaction mechanisms,
such as meson exchange currents and inelastic contributions, and
therefore cannot be used straightforwardly to extract $f_T$. Instead,
$f_T$ will be determined phenomenologically by fitting inclusive
quasielastic cross section data, following a procedure similar to that
used in the original SuSAM-v1 approach, as will be detailed later.

\paragraph{Longitudinal scaling function}

We
begin by analyzing the experimental data for the longitudinal scaling
function, which is obtained from $R_L$ by dividing out the appropriate
single-nucleon prefactor
\begin{equation}
f_L^*(\psi^*)_{\text exp} = \frac{R_L^{\text exp}}{Z r_{Lp} + N r_{Ln}}
\end{equation}

\begin{figure}[htp]
\centering
\includegraphics[scale=0.72, bb=108 545 434 778]{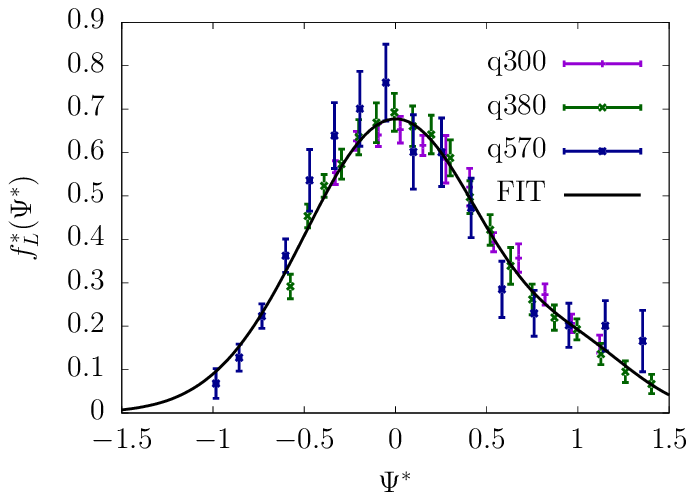}
\caption{Phenomenological longitudinal scaling function used in this work.
Data are for three values of
      \(q=300, 380, 570 \, \mathrm{MeV/c}\) from
      ref. \cite{Jou95,Jou96}.
}
\label{figura1}
\end{figure}

In Fig. \ref{figura1} we show the values of the longitudinal scaling
function $f_L^*(\psi^*)$ obtained from experimental data of $R_L$ for
${}^{12}\mathrm{C}$, scaled according to the procedure described
above. The figure illustrates that the longitudinal response exhibits
approximate scaling behavior when plotted as a function of
$\psi^*$. Although the scaling is not perfect, the deviations remain
within the experimental uncertainties, making it feasible to fit a
longitudinal scaling function $f_L^*(\psi^*)$ to all the
data points. Given the observed asymmetry in the shape of the
distribution, a sum of two Gaussian functions provides a suitable
phenomenological fit.
\begin{equation}
  f_L^*(\psi^*) =
  a_1 e^{-\frac{(\psi^* - b_1)^2}{2 c_1^2}} + a_2e^{-\frac{(\psi^* - b_2)^2}{2 c_2^2}}
\label{flstar}
\end{equation}
The  parameters $a_i$ (center), $b_i$ (width), and $c_i$ (height)
obtained for this model are presented in
Table \ref{tablerl}.
\begin{center}
\begin{table}[h]    
    \begin{ruledtabular}
    \begin{tabular}{lc}
        \textbf{Parameter} & \textbf{Value} \\
        \hline
        $a_1$ & 0.67732 \\
       $b_1$ & 0.0 \\
        $c_1$ & 0.497381 \\
        $a_2$ & 0.103387 \\
         $b_2$ & 1.05065  \\
        $c_2$ & 0.302541 \\
    \end{tabular}
    \end{ruledtabular}
\caption{Parameters for the phenomenological longitudinal scaling
  function described by the sum of two Gaussians.}
        \label{tablerl}
\end{table}
\end{center}

It is worth noting that the present longitudinal scaling function
$f_L^*(\psi^*)$ differs from the function $f_L(\psi)$ previously
extracted in Ref. \cite{Ama05}, which did not incorporate the
effective mass. In contrast, the new function $f_L^*(\psi^*)$ still
benefits from the dynamical relativistic effects associated with the
RMF model, as these are embedded through the use of the effective mass
in the scaling variable and the single-nucleon prefactors.

\paragraph{Transverse Scaling Function}

We now proceed to determine the new transverse scaling function
$f_T^*(\psi^*)$ by fitting quasielastic cross section data for
$^{12}$C after subtracting the longitudinal contribution. This
longitudinal part is computed using the phenomenological scaling
function $f_L^*(\psi^*)$ extracted in the previous step. We begin with
inclusive electron scattering data in the quasielastic region and, as
a first step, subtract the contribution from MEC in the two-particle
emission channel. This contribution is evaluated using a microscopic
model within the RMF framework \cite{Mar21a, Mar23b}. In the second
step, we subtract the longitudinal contribution already fixed by
$f_L^*(\psi^*)$. The remaining strength is assumed to be entirely
transverse in nature.
\begin{equation}
  (R_T)_{\text{exp}} =
  \frac{
    \left(\frac{d\sigma}{d\Omega d\omega}\right)_{\text{exp}}
    -\left(\frac{d\sigma}{d\Omega d\omega}\right)_{2p2h}
    -\sigma_M v_L R_L}{\sigma_M v_T}.
\label{rtexp}
\end{equation}
Dividing $(R_T)_{\text{exp}}$ by the single-nucleon transverse
response gives the experimental transverse
scaling function data:
\begin{equation}
(f_T^*)_{\text exp} = \frac{(R_T)_{\text{exp}}}{Z r_T^p + N r_T^n}.
\label{ftexp}
\end{equation}

\begin{figure}[htp]
\centering
\includegraphics[scale=0.6, bb=72 495 465 773]{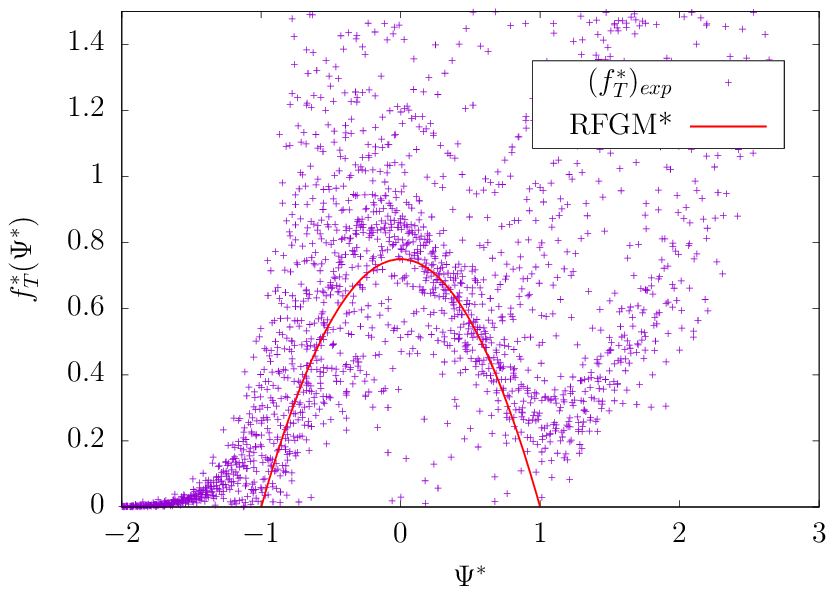}
\caption{
  Transverse scaling function data obtained from the
  cross section data after using Eqs. (\ref{rtexp}) and (\ref{ftexp}),
  compared with the Fermi Gas model scaling function
  using an effective mass \(M^* =0.8\).}
\label{figura2}
\end{figure}

Applying this procedure to the experimental $(e,e')$ cross section
data for $^{12}$C---consisting of approximately 3000 data points
available from the University of Virginia database
\cite{archive,archive2}---and representing the results as a function
of the scaling variable $\psi^*$, we obtain the distribution shown in
Fig. 2. It is apparent from the figure that the scaled data still
include significant contributions from inelastic nucleon excitations,
such as pion production and deep inelastic scattering. The next step
in the analysis is to remove these inelastic contributions by means of
a subtraction procedure.

\begin{figure*}[htp]
\centering
\includegraphics[width=15cm, bb= 2 32 541 729]{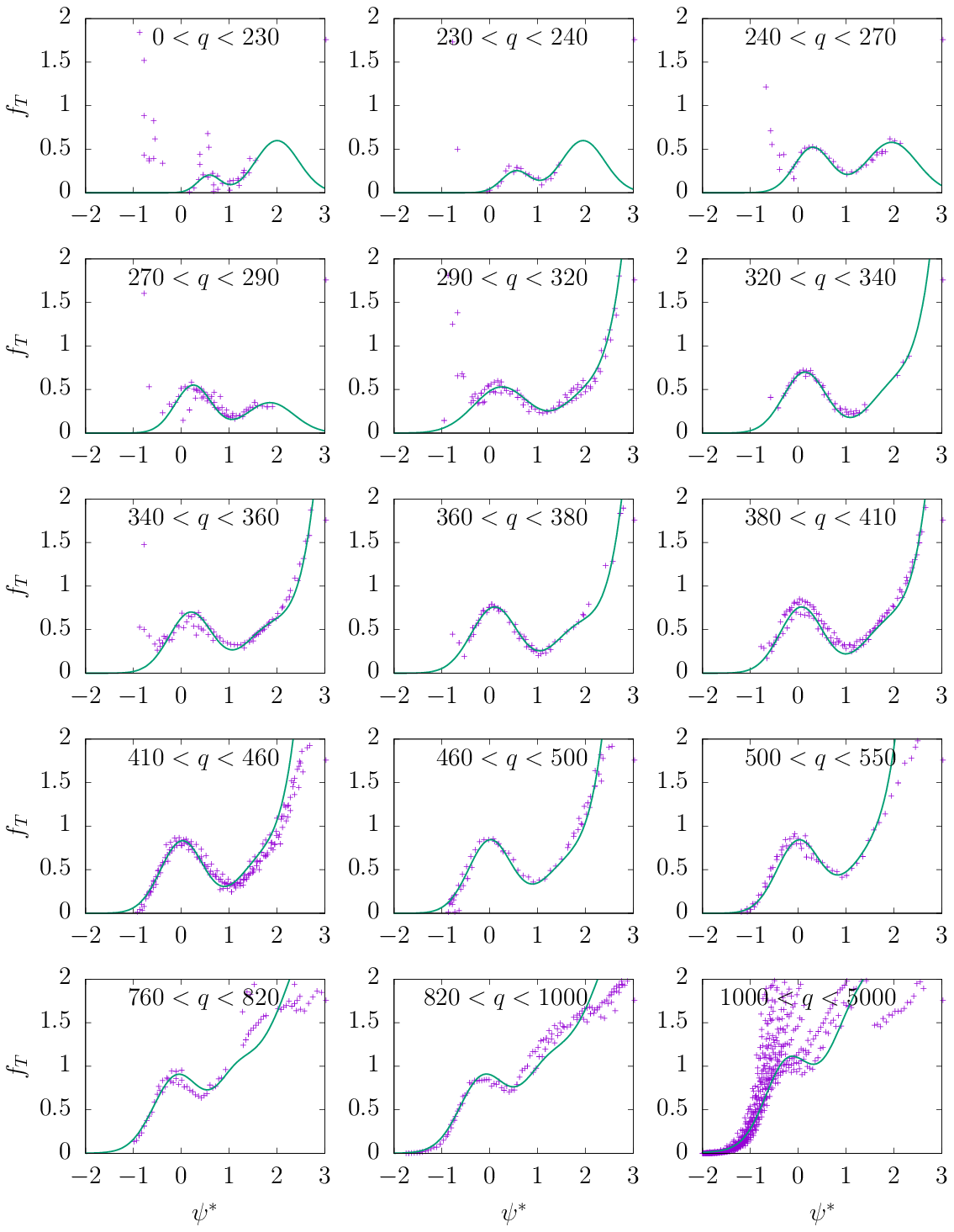}
\caption{ Experimental transverse scaling function data sets
  corresponding to different intervals of the momentum transfer.
  The green line represents a fit
  using two or three Gaussian functions.}
\label{figura3}
\end{figure*}

A careful analysis of the behavior of the experimental data as a function
of the transferred momentum $q$ indicates that the quasielastic
scaling function must carry an additional dependence on $q$, similar
to what was already implemented in the SuSA-v2 approach. This momentum
dependence cannot be ignored if one aims to accurately reproduce the
structure of the data across the full kinematic range.To account for this,
we performed independent fits by dividing the data into 18 intervals of 
$q$, as detailed in Table~\ref{tabla2}. A subset of 15 of these intervals
is displayed in separate panels in Fig.~\ref{figura3}, where we show the 
scaled experimental cross-section data grouped according to increasing 
$q$, from $q < 230$ MeV/c to $q > 1000$ MeV/c. The exact boundaries of
all intervals are provided in the table.

To extract the transverse scaling function in each momentum transfer
region, we fit the scaled cross section data using a sum of three
phenomenological Gaussian functions. The first Gaussian corresponds to
the quasielastic peak, the second describes the contribution from
the $\Delta(1232)$ resonance, and the third accounts for the rise
observed at higher energies due to deep inelastic scattering
(DIS). For momentum transfers below $q < 300 \, \mathrm{MeV}/c$, the
third Gaussian is omitted from the fit, as the experimental data show
no significant contribution from inelastic processes in this
regime. It is important to emphasize that this procedure does not rely
on a specific microscopic model for pion production or DIS; rather,
the region-by-region fitting allows us to determine the best
phenomenological parameters that describe each contribution based on
the data themselves. If more detailed models for the inelastic
channels become available, the fitting strategy could be refined
accordingly. However, such fine-tuning is beyond the scope of the
present approach, whose main objective is to capture the dominant
features of the quasielastic transverse response in a phenomenological and
model-independent manner.

The results of the region-by-region fits are also shown in
Fig. 3. Data corresponding to very high momentum transfers
($q>1000$ MeV/c), where deep inelastic contributions dominate, are fitted over a wide range of momentum transfer ($ 1000
< q < 5000
$) $\mathrm{MeV}/c$.  Additionally, for some low-$q$ intervals ($q < 360
\, \mathrm{MeV}/c$) and low energy, certain data points that clearly
do not exhibit scaling behavior were not taken into account for the fit. This selection
was performed by visually inspecting each momentum bin and discarding
points that deviate significantly from the quasielastic scaling
trend. After this data selection and fitting procedure, the first
Gaussian in each momentum transfer interval is identified as the
phenomenological transverse scaling function, defined independently
for each $q$-bin.
\begin{equation}
f_T^*(\psi^*,q) = a(q) e^{-\frac{(\psi^* - b(q))^2}{2 c(q)^2}}
\end{equation}

\begin{table}[ht]
\centering
\begin{ruledtabular}
\begin{tabular}{ccccc}
$q_i$ & $q_f$ & $a$ & $b$ & $c$ \\
\hline
0 & 230 & 0.20 & 0.58 & 0.24 \\
230 & 240 & 0.25 & 0.55 & 0.30 \\
240 & 270 & 0.52 & 0.30 & 0.38 \\
270 & 290 & 0.55 & 0.25 & 0.38 \\
290 & 320 & 0.53 & 0.25 & 0.58 \\
320 & 340 & 0.70 & 0.14 & 0.45 \\
340 & 360 & 0.70 & 0.20 & 0.45 \\
360 & 380 & 0.76 & 0.10 & 0.48 \\
380 & 410 & 0.76 & 0.07 & 0.45 \\
410 & 460 & 0.83 & 0.02 & 0.45 \\
460 & 500 & 0.84 & 0.01 & 0.45 \\
500 & 550 & 0.83 & 0.01 & 0.45 \\
550 & 610 & 0.84 & -0.01 & 0.46 \\
610 & 680 & 0.82 & -0.07 & 0.46 \\
680 & 760 & 0.83 & -0.09 & 0.46 \\
760 & 820 & 0.80 & -0.11 & 0.48 \\
820 & 1000 & 0.80 &-0.15 & 0.48 \\
1000 & 5000 & 0.83&-0.25 & 0.45 \\
\end{tabular}
\end{ruledtabular}
\caption{Parameters of the transverse scaling function fitted by a Gaussian with three parameters \(a\), \(b\), and \(c\) for different momentum transfer intervals.}
\label{tabla2}
\end{table}

\begin{figure}[htp]
\centering
\includegraphics[scale=0.6, bb=72 495 470 778]{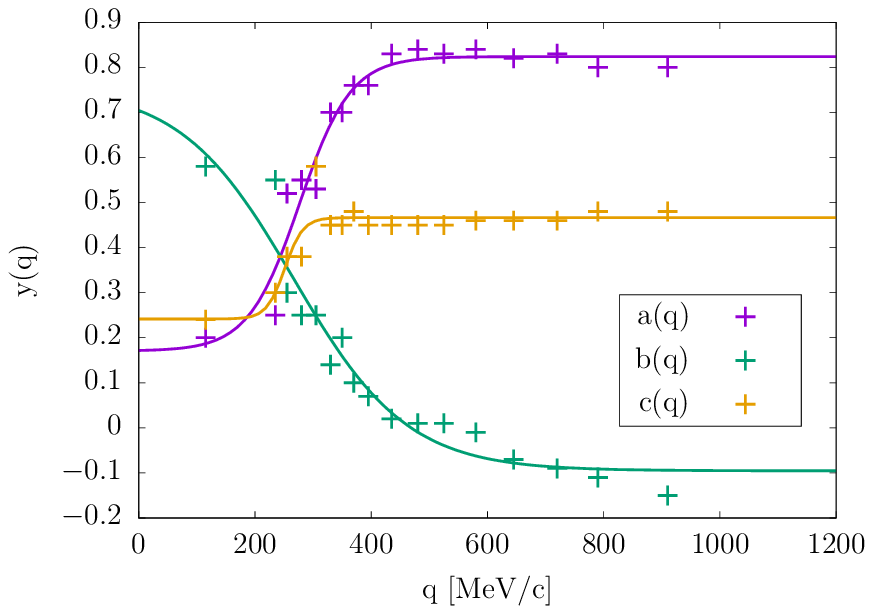}
\caption{ Coefficients \(a(q)\),
  \(b(q)\), and \(c(q)\) of the transverse scaling function as a
  function of the momentum transfer, fitted using a Fermi function.
}
\label{figura35}
\end{figure}

The parameters of the transverse scaling function $f_T$, denoted as
$a(q)$, $b(q)$, and $c(q)$, are listed in Table~\ref{tabla2} for each momentum
transfer interval. These parameters exhibit a clear dependence on
$q$. In particular, the position of the peak, given by $b(q)$, shows a
noticeable shift: it starts around $b \sim 0.5$ for low $q$ values
(below 240 MeV/$c$), then decreases rapidly, crossing zero near $q =
500 \, \mathrm{MeV}/c$, and becomes increasingly negative, reaching
approximately $b = -0.25$ at $q = 1000 \, \mathrm{MeV}/c$. Similarly,
the height of the scaling function, $a(q)$, increases with $q$ up to
around 400 MeV/$c$, where it stabilizes. The width parameter $c(q)$
follows a similar trend, stabilizing for $q \gtrsim 300 \,
\mathrm{MeV}/c$. This explicit momentum-transfer dependence is
essential to accurately reproduce the behavior of the cross section in
the low-$q$ region, where violations of scaling are more
pronounced. Therefore, the inclusion of this $q$-dependence in the
updated SuSAM-v2 model is expected to significantly improve the
description of the quasielastic region at low momentum transfers,
which could not be adequately captured using a scaling function
independent of $q$.

This behavior of the parameters is more clearly illustrated in
Fig. \ref{figura35}, where they are plotted as a function of the
momentum transfer $q$. It is found that this dependence can be
accurately described using a single Fermi function for each parameter,
\begin{equation}
  y(q)=    \frac{\alpha_1}{1 + e^{(q - \alpha_2) / \alpha_3}} + \alpha_4
\end{equation}
where $y= a,b,c$.
This provides a smooth
parametrization of the transverse scaling function $f_T^*(\psi^*,
q)$.
The corresponding fitted values are listed in
Table~\ref{tabla3}. The use of Fermi functions ensures a reliable
extrapolation of the scaling function across a wide range of momentum
transfers.

\begin{table}[ht]
\centering
\begin{ruledtabular}
\begin{tabular}{cccc}
   Coefficient & $a(q)$ & $b(q)$  & $c(q)$\\  
\hline
 $\alpha_1$           & $-0.653271$ & 0.851301     & $-0.224921$ \\ 
 $\alpha_2$ [MeV/c]  & 276.616     & 266.775      & 251.325\\ 
 $\alpha_3$ [MeV/c]  &  44.0707    & 97.4585      &  15.5872 \\ 
 $\alpha_4$          &  0.823775   & $-0.0952112$ & 0.466494\\ 
        \hline
\end{tabular}
\end{ruledtabular}
\caption{Parameters of the Fermi function that fit the coefficients
  \(a\), \(b\), and \(c\) of the phenomenological transverse scaling
  function.}
\label{tabla3}
\end{table}

\begin{figure}[htp]
\centering
\includegraphics[width=8.2cm,bb=33 265 517 765]{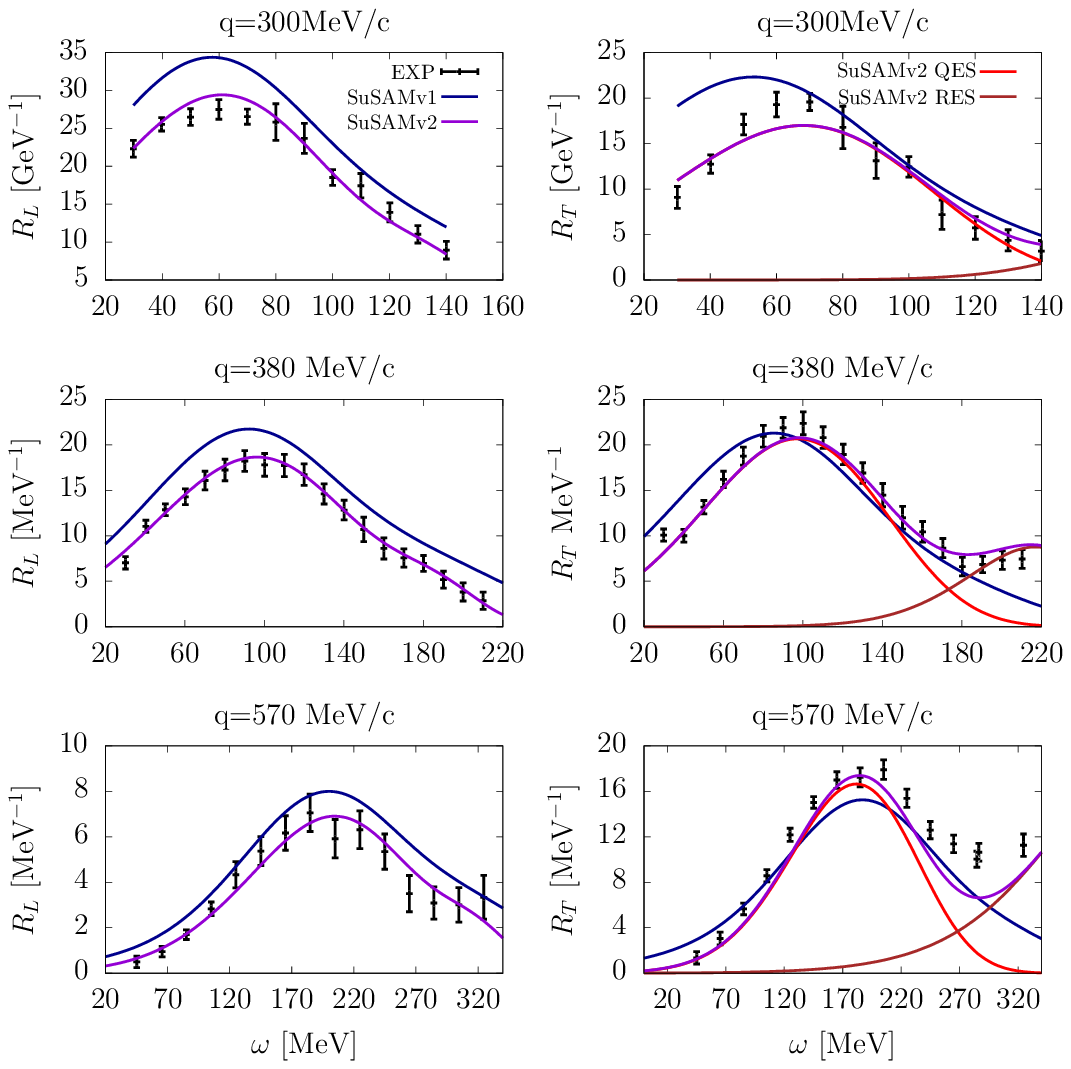}
\caption{Separated longitudinal and transverse electromagnetic
  responses for three momentum transfers: \(q = 300\), 380, and 570
  MeV/\emph{c}, using the SuSAM-v1 and SuSAM-v2
  prescriptions. The results are compared with experimental data from
  Ref.~\cite{Jou95,Jou96}.
  }
\label{jourdan}
\end{figure}

\added{The $q$-dependence observed in the extracted transverse scaling
function $f_T^*$ may be, at least in part, attributed to Final State Interactions (FSI).
These effects are known to induce violations of scaling, particularly
at low and high momentum transfer, due to mechanisms such as nucleon
rescattering and absorption \cite{Arr99,Ben99}. Although FSI are not
explicitly included in the present model, their effects are effectively
incorporated through the phenomenological $q$-dependence of the scaling function.}

\section{Results}
\label{seccion4}

Once the new scaling analysis SuSAM-v2 of the electron scattering data has been
completed, and the two phenomenological scaling functions
$f_L^*(\psi^*)$ and $f_T^*(\psi^*, q)$ have been determined, we
proceed to analyze how the model describes the available data for
response functions and inclusive cross sections. This includes both
electron scattering and predictions for neutrino-induced
reactions. The main objective is to assess whether the SuSAM-v2 model
provides an improved description compared to the previous SuSAM-v1
version. A positive outcome would reinforce the reliability of the new
approach and have important implications for the modeling of
quasielastic lepton-nucleus interactions, particularly in the context
of neutrino oscillation experiments.

\subsection{Electron scattering}

The SuSAM-v2 model preserves the same factorized structure used
in the previous version, where the response functions are expressed as
the product of a single-nucleon term and a phenomenological scaling
function. The most significant difference now is that the longitudinal and
transverse responses are computed separately using their respective
phenomenological scaling functions, $f_L^*(\psi^*)$ and $f_T^*(\psi^*,
q)$, obtained from the independent analysis of longitudinal and
transverse electron scattering data. This separation allows for a more
flexible and accurate description of each response channel
\begin{eqnarray}
  R_L(q,\omega) & =& (Z  r^p_L+N r_L^n)  f_L^*(\psi^*), \\
  R_T(q,\omega) & =& (Z  r^p_T+N r_T^n)  f_T^*(\psi^*,q).
\end{eqnarray}

\begin{figure*}[htp]
\centering
\includegraphics[width=13cm,bb=12 259 534 766]{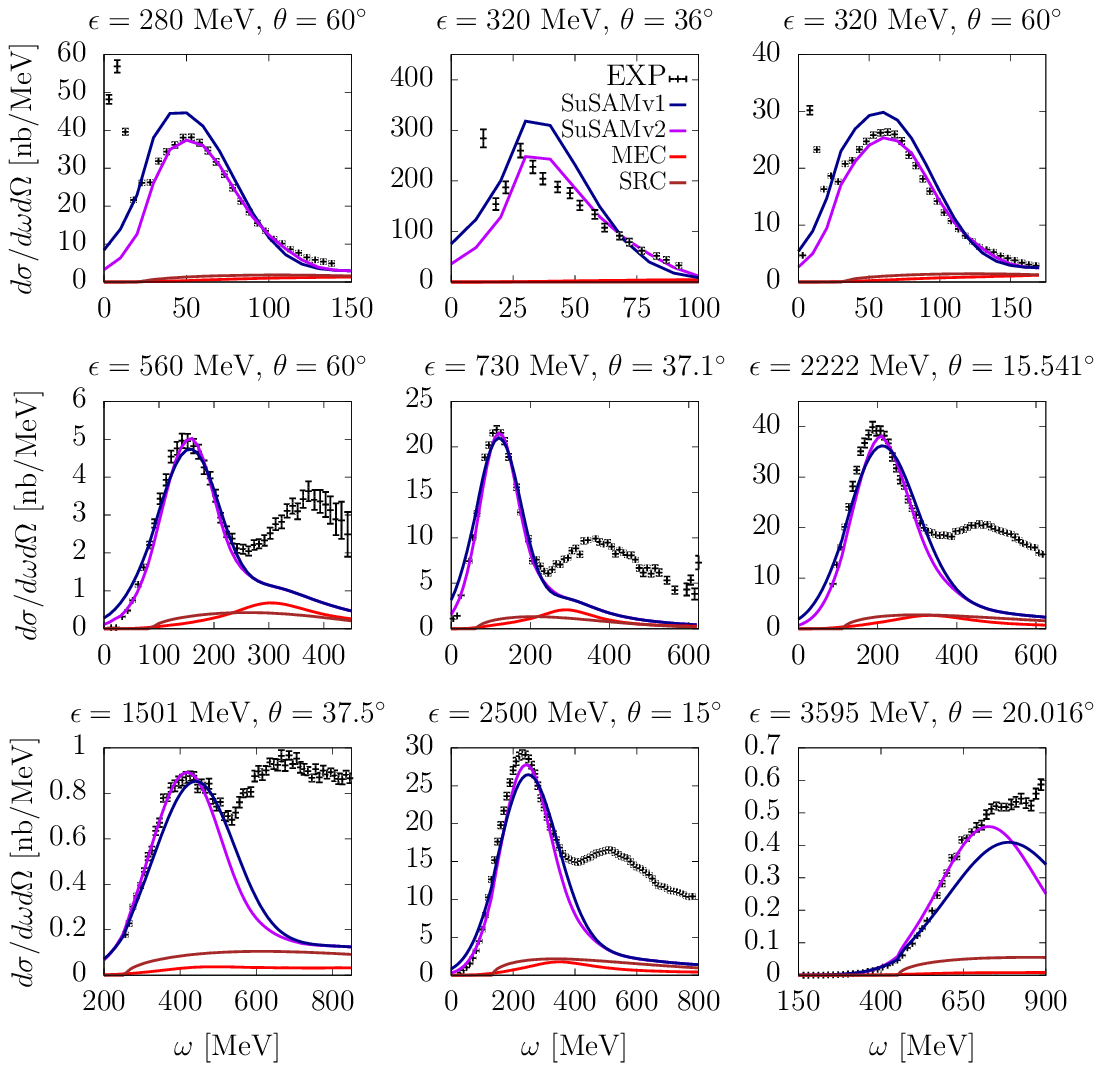}
\caption{Inclusive electron scattering cross section on \(^{12}\)C for
  nine different kinematics. Comparison between low (top panels),
  intermediate (middle panels), and high (bottom panels) momentum
  transfer using the SuSAM-v1 and SuSAM-v2 models, including the
  2p2h contribution from MEC and an estimation from SRC.
  The experimental data are taken from
  Refs.~\cite{archive,archive2,Ben08}.}
\label{electrones}
\end{figure*}

The quasielastic longitudinal and transverse response functions for
inclusive $(e,e')$ scattering from $^{12}$C are shown in
Figure~\ref{jourdan}. Alongside the experimental data, we present the
theoretical results obtained with the updated SuSAM-v2 model and
compare them with those from the previous SuSAM-v1 version. For the
transverse response, we also include the contribution from the
resonant component that we have substracted (parametrized as a sum of
Gaussians), that contributes to the tail of the $T$-response for high $\omega$.

The first observation is that the longitudinal response $R_L$ is well
reproduced by the SuSAM-v2 model for all three momentum transfers
shown. This result is expected, since the $R_L$ data were used
directly in the extraction of the scaling function $f_L^*$. In
contrast, the SuSAM-v1 model clearly overestimates $R_L$, highlighting
an initial improvement introduced by the new version. A significant
improvement is also observed in the description of the transverse
response $R_T$ with SuSAM-v2. This constitutes a genuine prediction of
the model, since $R_T$ data were not used in the fit of the transverse
scaling function $f_T^*$—only the inclusive cross section data were
employed. This outcome indicates that a simultaneous fit to $R_L$ and
to the inclusive cross section, within our analysis framework, leads
naturally to a reliable description of $R_T$. In particular, for $q =
570$ MeV/c, the new model incorporates the necessary enhancement
around the quasielastic peak to reproduce the observed data, which was
underestimated in the previous version.
It is  worth highlighting the improved
agreement achieved at low momentum transfer, where scaling violations
are more pronounced and the previous model fails to reproduce both the
shape and magnitude of the response.

In Fig. \ref{electrones} we show the inclusive electron scattering
cross section on $^{12}$C for nine different kinematics, covering low
(top panels), intermediate (middle panels), and high (bottom panels)
momentum transfers. The results obtained with the updated
SuSAM-v2 model are compared with those of the previous version,
SuSAM-v1, and with experimental data.

The comparison shows that the SuSAM-v1 model tends to
overestimate the cross section at low momentum transfers, particularly
around the quasielastic peak. At higher values of $q$, it often
underestimates the peak height and, in some cases, fails to reproduce
the correct peak position. In contrast, the new SuSAM-v2 model
exhibits a clear improvement in all these aspects, yielding a more
accurate description of both the magnitude and the location of the
quasielastic peak across all kinematic regimes.

A noticeable enhancement is also observed at low energy transfer,
especially at low momentum transfer, where the new model better
reproduces the behavior of the data. However, it is worth noting that
the description of the region corresponding to giant resonances
remains poor, as these data were not included in the fitting procedure
and lie outside the quasielastic domain targeted by our model.

Regarding the 2p2h contribution, taken from refs. \cite{Mar21a,Mar23b},
it is based in a microscopic calculation of meson-exchange currents in
the RMF of nuclear matter and an estimation of 2p2h emission by the
one-body current  describing the high energy tail of the scaling
function.
In the figure the 2p2h contributon allows to estimate its order of magnitude
and its potential role in improving the description of the dip region
between the quasielastic peak and the onset of pion production. Since
our model is designed specifically to describe the quasielastic
region, pion production mechanisms are not included. Nonetheless, the
inclusion of 2p2h provides a reasonable approximation of the expected
enhancement in this transition region.

\subsection{Neutrino scattering}

We now proceed to the results section for neutrino-induced
reactions. In the case of the $(\nu_{\mu},\mu^-)$ process, the
double-differential cross section is given by
\begin{eqnarray}
\frac{d\sigma}{dT_\mu d\cos\theta_\mu}
&=&
\sigma_0 \left( v_{CC} R_{CC}+2 v_{CL} R_{CL}+v_{LL} R_{LL} \right.
\nonumber\\
&&\mbox{}\left.
+v_{T} R_{T}
\pm 2v_{T'} R_{T'} \right),
\end{eqnarray}
where $\theta_\mu$ and $T_\mu$ are muon scattering angle and kinetic
energy, respectively.  Here $\sigma_{0}$ is the analogue of the Mott
cross section for neutrino scattering, the coefficients $v_K$ are
purely kinematical factors, and five nuclear response functions
$R_K(q,\omega)$ contribute to the cross section (the $\pm$ sign
corresponds to neutrino or antineutrino scattering). Detailed
definitions of these quantities can be found in Ref. \cite{Rui18}.

\begin{figure*}[htp]
\centering
\includegraphics[width=13cm,bb=21 241 523 663]{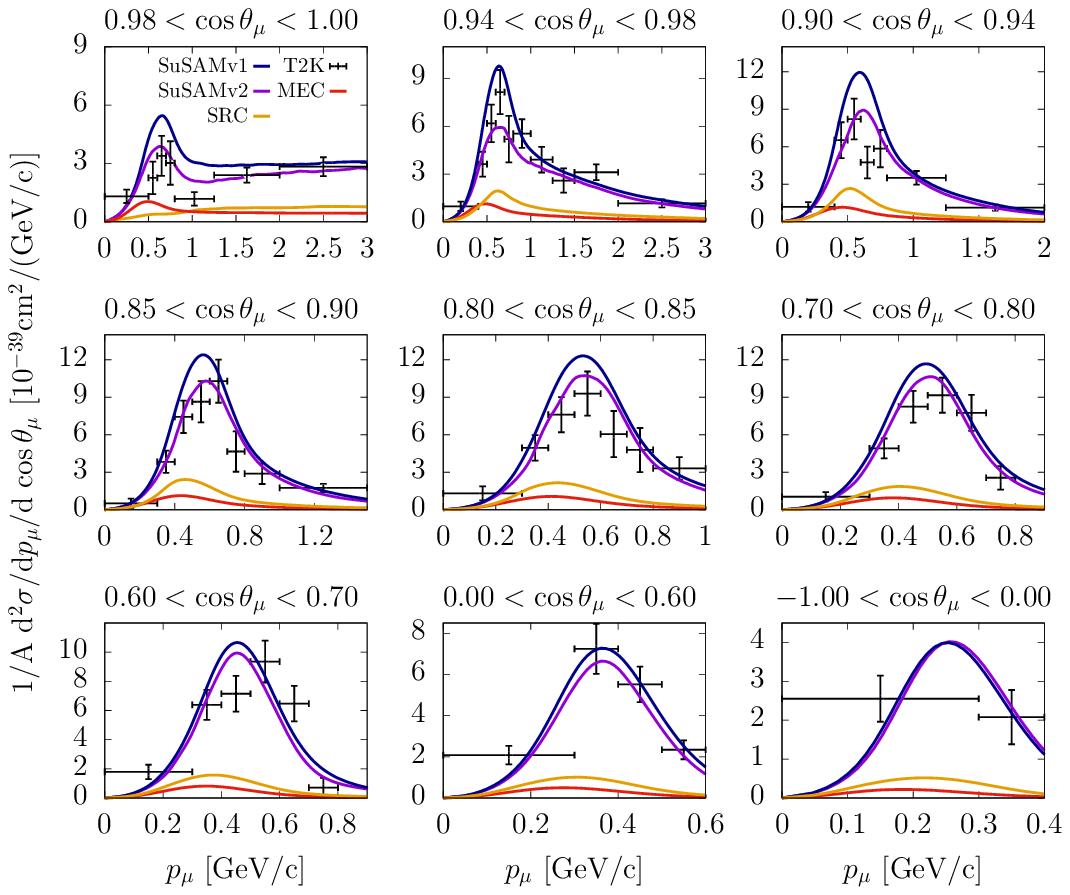}
\caption{
  Neutrino charged-current quasi-elastic cross section from
  the T2K experiment, shown as a flux-folded double-differential
  distribution per nucleon. The comparison has been performed using
  the SuSAM-v1 and SuSAM-v2 models, including the 2p2h
  contribution. The experimental data, taken on a \(^{12}\)C target,
  are from Ref. \cite{Abe16}. }
\label{t2k}
\end{figure*}

The SuSAM model assumes that each nuclear response can be factorized
into a single-nucleon prefactor multiplied by a universal scaling
function. In the original SuSAM-v1 approach, this scaling function was
extracted from inclusive electron scattering data and was assumed to
be the same for all channels. In the present work, within the SuSAM-v2
framework, we retain the same factorization scheme but distinguish
between longitudinal and transverse scaling functions, denoted
$f_L^*(\psi^*)$ and $f_T^*(\psi^*,q)$, respectively. Specifically, we
write:
\begin{eqnarray*}
R_K(q,\omega) & = & N r_Kf_L^*(\psi^*), \kern 1cm K=CC,CL,LL, \\
R_K(q,\omega) & = & N r_K f_T^*(\psi^*,q), \kern 1cm K=T,T',
\end{eqnarray*}
where $r_K$ are
the single-nucleon prefactors. In this
approach, it is further assumed that the same transverse scaling
function applies to both the $T$ and $T'$ channels, i.e., $f_T^*$ is
used for both $R_T$ and $R_{T'}$.

One essential difference in neutrino experiments is that the measured
cross sections are flux-averaged, meaning they are weighted integrals
over the incoming neutrino energy spectrum. Since each experiment has
a different neutrino flux, the measured observables effectively emphasize
different energy regions of the cross section. This must be taken into
account when comparing theory and experiment.  The flux-averaged
double differential cross section is computed as
\begin{equation}
\frac{d^2\sigma}{dT_\mu d\cos\theta_\mu}
=
\frac{
\int dE_\nu \Phi(E_\nu)
\frac{d^2\sigma
}{dT_\mu d\cos\theta_\mu}(E_\nu)}
{\int dE_\nu \Phi(E_\nu)}, 
\label{average}
\end{equation}
where $\Phi(E_\nu)$ is the neutrino flux
and
$\frac{d^2\sigma
}{dT_\mu d\cos\theta_\mu}(E_\nu)$
is the cross section for fixed neutrino energy $E_\nu$.

In this section, we present results for two representative
experiments: T2K, which uses a $^{12}$C target, and MINER$\nu$A, which
uses a hydrocarbon (CH) target. In each case, we compare the
predictions of the SuSAM v1 and v2 models, showing the
effects of the updated scaling functions. A model for the 2p2h contribution is
also included additively, as it represents an independent reaction
channel not captured by the quasielastic scaling functions.

The 2p2h channel includes two separate contributions. The first arises 
from meson-exchange currents, for which we employ a fully microscopic 
model as described in Ref.\cite{Mar23b} for the inclusive case and in 
Refs.\cite{Mar24a,Mar24b} for the semi-inclusive cross section. In this
approach, the 2p2h-MEC response functions are calculated using the 
complete $\Delta$-propagator, with in-medium effects incorporated through 
an effective mass and a vector energy for the $\Delta$ resonance. The 
calculation is performed under the universal coupling approximation.

The second 2p2h contribution is related to the one-body current and
is linked to the emission of correlated nucleon pairs. This term is
described using a semi-empirical formula introduced in
Refs. \cite{Mar23a,Mar23b}, which is proportional to the 2p2h phase
space and the nucleon form factors, with fitted coefficients designed
to reproduce the high-energy tail of the experimental scaling
function. This contribution is
phenomenological in nature and is intended as a placeholder for the
effects of nucleon-nucleon correlations, in the absence of a complete
microscopic model.

We begin by presenting results for the T2K experiment in Fig. 7. The
quasielastic cross section is reported in bins of the outgoing muon
angle $\cos\theta_\mu$, which are not equally spaced. As a result,
an average over each bin width is required to compute the
theoretical prediction. Each panel of Fig. 7
displays the cross section as a function of the muon momentum for a
given angular bin, comparing the SuSAM-v1 and SuSAM-v2
predictions. The total prediction includes the 2p2h contributions from
both MEC and SRC, and we also show these components separately for
reference.

\begin{figure*}[htp]
\centering
\includegraphics[width=13cm,bb=3 230 547 718]{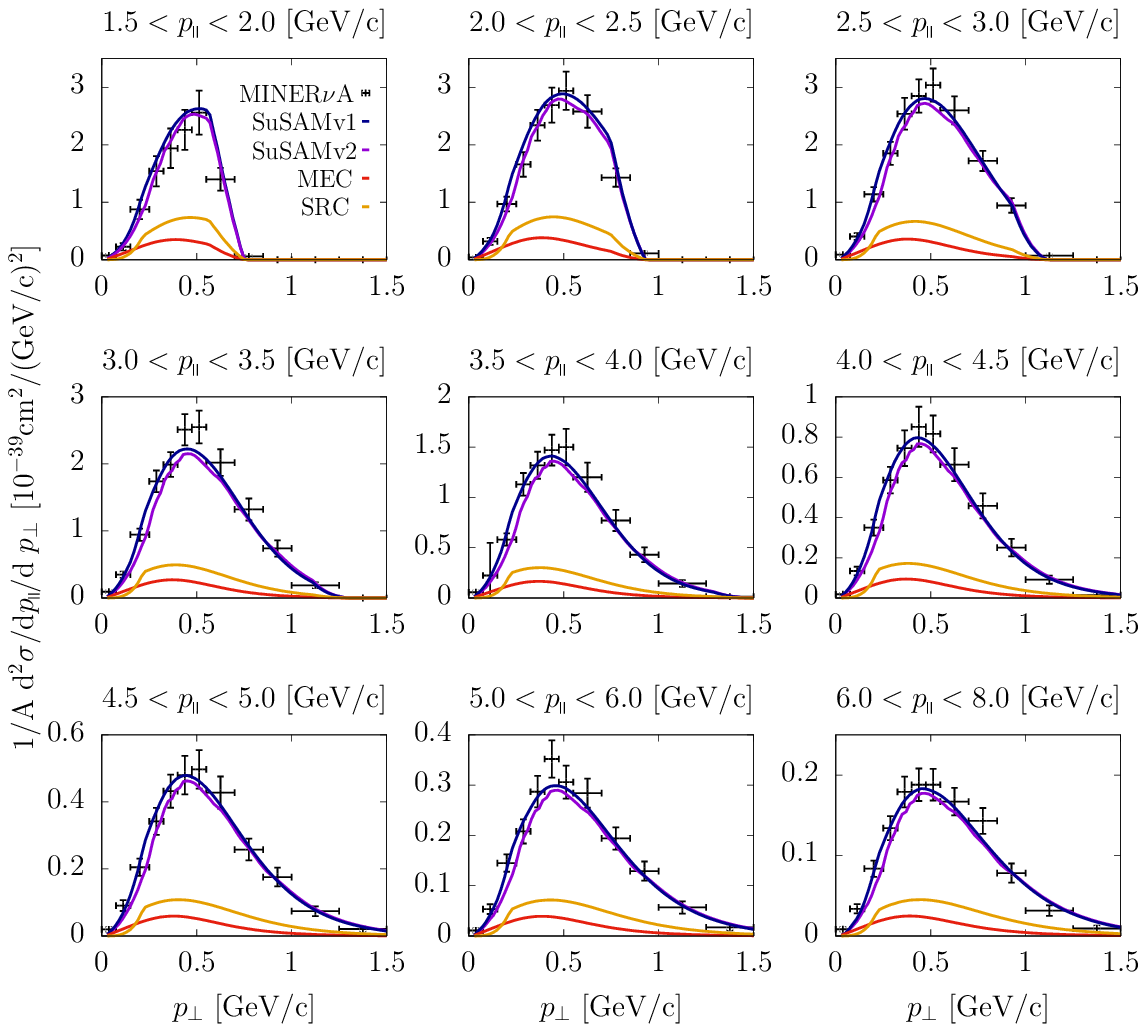}
\caption{
  Neutrino charged-current cross section from the MINER$\nu$A
  experiment, shown in bins of transverse muon momentum and as a
  function of parallel muon momentum. The comparison has been
  performed using the SuSAM-v1 and SuSAM-v2 models, including
  the 2p2h contribution. The experimental data, taken on CH target,
  are from Ref. \cite{Rut19}.}
\label{minerva}
\end{figure*}

The analysis of Fig. 7 reveals that the SuSAM-v2 results are generally
reduced compared to those of SuSAM$^*$v1, leading to an improved
agreement with the experimental measurements. This improvement is
especially significant at small muon scattering angles, where the v1
version tended to overestimate the cross section near the peak. The
reduction in SuSAM-v2 arises primarily from two modifications. First,
the longitudinal scaling function has been fitted, resulting in a
slight decrease of the longitudinal response. Second, the transverse
scaling function now depends explicitly on the momentum transfer $q$;
for low $q$, its height and width are significantly reduced, and its
peak position is shifted. These changes contribute to a more accurate
description of the cross section at small angles, where low momentum
transfers dominate the response, thus enhancing the overall
consistency with the T2K data.  The effects of MEC and SRC show
similar features, with some variations at very small angles. Both
responses peak at lower energies compared to the 1p1h response. The
inclusion of 2p2h contributions from both MEC and correlations is
important for accurately describing the data, accounting for roughly
20\% of the total cross section.

To conclude this work, we compare the model predictions with the
double-differential cross sections measured by the MINER$\nu$A experiment
on a CH target. The MINER$\nu$A data provide a stringent test for
theoretical models, as they span a wide kinematic range with neutrino
energies extending well beyond 1.5 GeV and cover longitudinal muon
momenta up to 20 GeV. This implies that the corresponding momentum
transfers probe regions significantly higher than those accessed in
lower-energy experiments such as T2K, which is largely limited to
momentum transfers below 1 GeV/c. The cross sections are reported as
functions of the muon momentum components parallel and perpendicular
to the incoming neutrino direction, variables which are particularly
sensitive to nuclear effects and allow for a more differential
analysis of the reaction dynamics. An important feature of MINER$\nu$A is
its ability to isolate the quasielastic-like contribution from other
channels, enabling a focused comparison with theoretical predictions
in the QE region. Moreover, these data provide valuable insight into
the role of multinucleon emission mechanisms: our analysis includes
both meson-exchange currents and short-range correlation 
contributions in the 2p2h sector, which turn out to be essential for a
consistent description of the measured cross sections.

Figure 8 presents the comparison between the SuSAM model predictions
and the double-differential cross sections measured by MINER$\nu$A as
functions of the transverse muon momentum $p_T$, for fixed bins of
longitudinal momentum $p_\parallel$ ranging from 1.5 GeV to 8 GeV. The
predictions from both SuSAM-v1 and the updated SuSAM-v2 models are
shown. In all cases, the theoretical cross sections include both the
quasielastic contribution and the 2p2h
components.
As observed in the figure, both versions of the model provide a very
good description of the experimental data across the full kinematic
range. The agreement holds for low and high $p_\parallel$ bins, with
both SuSAM-v1 and SuSAM-v2 reproducing the shape and magnitude of the
cross sections. The overall differences between the two versions are
small. The most noticeable difference appears in the width of the
distribution as a function of $p_T$: the SuSAM-v2 prediction is
slightly narrower, especially in the lower $p_\parallel$ bins. This
reflects the impact of the updated scaling functions $f_L^*(\psi^*)$
and $f_T^*(\psi^*,q)$ introduced in SuSAM-v2, which were extracted
independently from longitudinal and transverse electron scattering
data, respectively.

The similarity between the two models in this particular data set can
be attributed to the neutrino flux of MINER$\nu$A, which peaks at
relatively high energies. At large momentum transfers, where both
models yield similar scaling behavior and the longitudinal and
transverse responses converge, the differences between SuSAM-v1 and
SuSAM-v2 become less significant. 

Another important feature of the comparison is the size of the 2p2h
contribution. In all kinematic bins, the inclusion of the 2p2h
response—driven by MEC and SRC—leads to a visible enhancement of the
total cross section. The contribution of 2p2h processes is
non-negligible, amounting to approximately 30\% of the total cross
section in most bins. This confirms the relevance of multinucleon
effects in the interpretation of inclusive neutrino–nucleus data and
underlines the need for their consistent incorporation in theoretical
models, especially for oscillation experiments.

While SuSAM-v1 already provided a satisfactory description of the
MINER$\nu$A data, the updated SuSAM-v2 model reinforces these results by
offering a more flexible and accurate framework. The independent
treatment of the longitudinal and transverse nuclear responses
improves the predictive power of the model, particularly at lower
energies and in situations where the separation of response functions
becomes essential. Although the differences between v1 and v2 are
minor in the case of MINER$\nu$A, SuSAM-v2 is expected to provide
significant improvements in other kinematical regions and for other
observables where SuSAM-v1 is less reliable.

\begin{figure}[htp]
\centering
\includegraphics[width=8.5cm,bb=66 440 463 722]{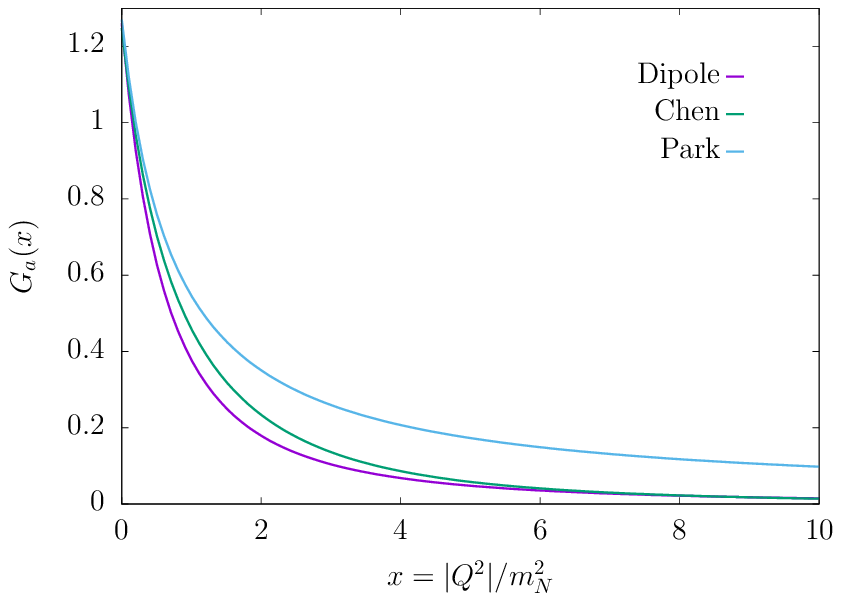}
\caption{Axial form factor of the nucleon as a function of $x=|Q^2|/m_N^2$, comparing the traditional dipole form with recent parametrizations from \cite{Par21,Che21,Che22}.}
\label{ga}
\end{figure}

\begin{figure*}[htp]
\centering
\includegraphics[width=13cm,bb=1 372 547 734]{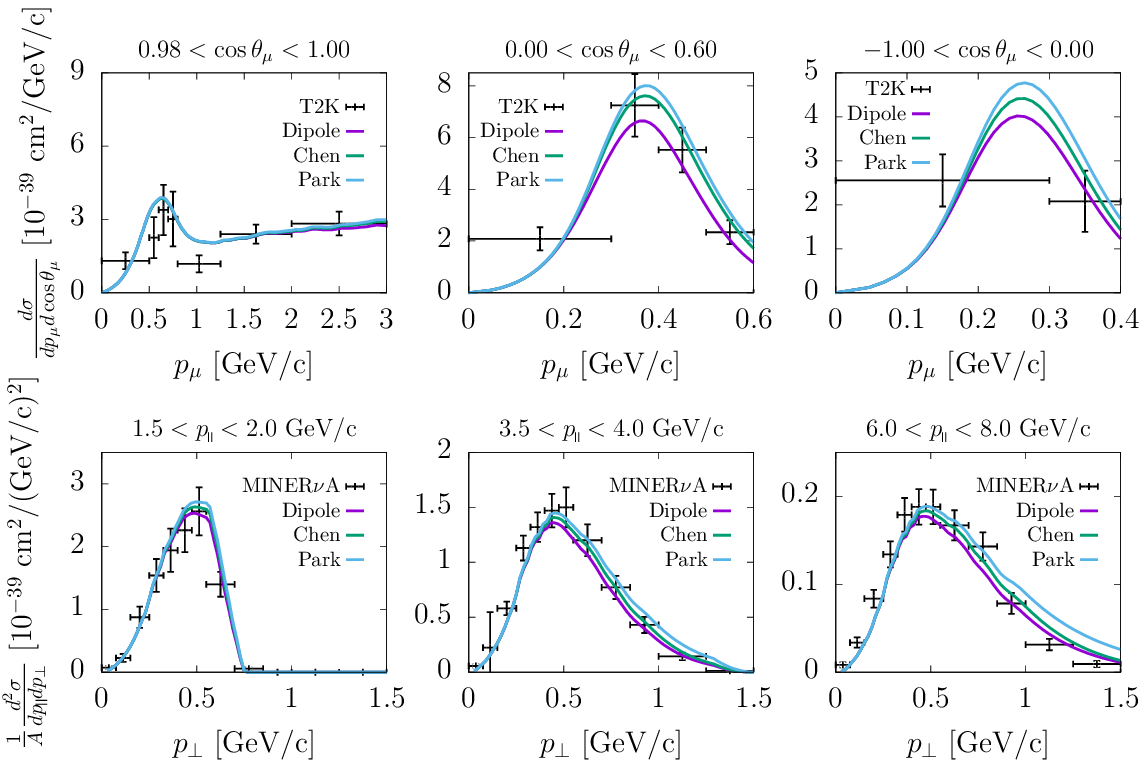}
\caption{Impact of different axial form factor parametrizations on the predicted cross sections. The calculations use dipole (standard), Chen et al.\cite{Che21,Che22}, and Prat et al.\cite{Par21}\ parametrizations for selected T2K and MINERvA kinematics.}
\label{neuaxial}
\end{figure*}

\added{The axial form factor of the nucleon is commonly modeled
using the standard dipole parametrization with $G_A(0) = 1.26$ 
and an axial mass $M_A = 1.032$~GeV. This choice ensures consistency 
with previous SuSAM-based analyses. However, recent experimental 
and lattice QCD studies~\cite{Par21,Che21,Che22} indicate noticeable
deviations from the dipole shape, particularly at intermediate momentum
transfers, as shown in Fig. \ref{ga}. These findings suggest that more refined descriptions of
$G_A$ may be needed to reduce systematic uncertainties in neutrino-nucleus 
scattering models. A full analysis of these effects is beyond the scope of 
the present work; nevertheless, to illustrate their potential impact, in 
Fig.~\ref{neuaxial} we compare the double-differential cross sections obtained
using different axial form factor parametrizations (namely, those of Dipole,
Chen and Park) for selected kinematic configurations relevant
to T2K and MINER$\nu$A. The results show that the differences between 
parametrizations can reach non-negligible levels in specific regions 
of phase space, in some cases increasing the predicted cross section 
by up to 25\% at the QE-peak. This emphasizes the importance of a more 
accurate treatment of the axial structure in precision neutrino physics.}

\section{Conclusions}
\label{seccion5}

In this work, we have developed and tested a new version of the SuSAM
model—SuSAM-v2—for describing quasielastic lepton–nucleus scattering.
The model is based on the scaling
analysis of inclusive electron scattering data, which allows the
nuclear responses to be factorized into a single-nucleon cross section
and a universal scaling function. The key improvement introduced in
SuSAM-v2 is the separate treatment of the longitudinal and transverse
scaling functions, $f_L^*(\psi^*)$ and $f_T^*(\psi^*, q)$, both
extracted from electron scattering data and fitted
independently. These functions are parametrized with simple analytical
forms, enabling an efficient and transparent implementation of the
model across a wide kinematic range.
The transverse scaling function in SuSAM-v2 is extracted independently
for each momentum transfer interval using a model-independent
procedure, where the contributions from pion production and deep
inelastic scattering are subtracted.  The resonance region is not
described in detail but rather approximated in a coarse manner, with
the sole purpose of identifying and isolating the quasielastic domain;
the rest of the inelastic spectrum is disregarded in the scaling
analysis.

Compared to the earlier SuSAM-v1, which employed a common scaling
function for all nuclear responses, SuSAM-v2 provides an improved
description of the separated longitudinal and transverse
electromagnetic response functions, particularly at intermediate and low
momentum transfers. This enhancement translates directly into a more
accurate and flexible framework for predicting neutrino–nucleus cross
sections in the quasielastic channel. We have validated the model by
comparing its predictions with $^{12}$C$(e,e')$ cross sections, and
T2K and MINER$\nu$A data for CC muon neutrino
scattering, showing excellent agreement in both shape
and magnitude of the double-differential cross sections, including the
contributions from 2p2h  excitations, computed with a semiempirical model.

All superscaling models---including
SuSAM---are ultimately phenomenological parametrizations of the
inclusive cross section, each incorporating different nuclear
ingredients. In the case of SuSAM, the effective mass of the nucleon
in the medium plays a central role in shaping the scaling variable and
the nuclear dynamics. This contrasts with the widely used SuSA model,
which does not include an effective mass but requires the use of an
empirical separation energy to shift the scaling function. Despite
these differences, both models reproduce inclusive electron and
neutrino scattering data with comparable accuracy.

Notably, both SuSAM-v2 and SuSA-v2 incorporate a transverse scaling
function that depends on the momentum transfer $q$.
However, while SuSA-v2
derives this function from a RMF model for
finite nuclei—resulting in a numerically defined, non-analytic
function—SuSAM-v2 provides an explicit analytical parametrization
based on a Gaussian form, whose parameters vary smoothly with
$q$. This feature makes SuSAM-v2 particularly well-suited for use in
Monte Carlo event generators and phenomenological studies, where
reproducibility and computational efficiency are essential.
Scaling-based parametrizations such as SuSAM-v2 provide a
valuable alternative to fully microscopic approaches, offering both
physical transparency and practical advantages for interpreting
experimental data and supporting neutrino oscillation analyses.
Multiple models based on scaling approaches
offers the possibility to estimate the theoretical or model-dependent
systematic uncertainties that affect neutrino cross section analyses
and oscillation experiments. As
neutrino experiments continue to improve in precision,
phenomenological models with solid empirical foundations and simple
implementations will remain a key tool in understanding the complex
dynamics of neutrino–nucleus interactions.

\begin{acknowledgments}

We thank Jesus Gonzalez-Rosa and Jorge Segovia for helpful discussions.

\vskip 0.2cm

V.L.M.C.  acknowledges financial support provided by Ministerio
Español de Ciencia, Innovación y Universidades under grant
No. PID2022-140440NB-C22; Junta de Andalucía under contract Nos. PAIDI
FQM-370 and PCI+D+i under the title: ``Tecnologías avanzadas para la
exploración del universo y sus componentes'' (Code AST22-0001).

The work was supported by Grant No. PID2023-147072NB-I00 funded by
MICIU/AEI /10.13039/501100011033 and by ERDF/EU; by Grant No. FQM-225
funded by Junta de Andalucia.

\end{acknowledgments}


\end{document}